\documentclass[twocolumn]{revtex4}
\usepackage{amssymb,amsmath}
\usepackage{graphicx}
\usepackage{dcolumn}

\begin{document}

\title{\textbf{\fontfamily{phv}\selectfont Single-electron quantum dot in Si/SiGe with integrated charge-sensing}}
\author{C. B. Simmons, Madhu Thalakulam, Nakul Shaji, Levente J. Klein, Hua Qin,\\ R. H. Blick, D. E. Savage, M. G. Lagally, S. N. Coppersmith, M. A. Eriksson\\University of Wisconsin-Madison, Madison, WI 53706}

\begin{abstract}
Single-electron occupation is an essential component to measurement and manipulation of spin in quantum dots, capabilities that are important for quantum information processing.  Si/SiGe is of interest for semiconductor spin qubits, but single-electron quantum dots have not yet been achieved in this system.  We report the fabrication and measurement of a top-gated quantum dot occupied by a single electron in a Si/SiGe heterostructure.  Transport through the quantum dot is directly correlated with charge-sensing from an integrated quantum point contact, and this charge-sensing is used to confirm single-electron occupancy in the quantum dot.
\end{abstract}

\maketitle

Semiconductor quantum dots provide highly tunable structures for trapping and manipulating individual electrons,$^{1, 2}$ with significant potential for integration and scaling, and therefore are promising candidates as qubits for quantum computation.$^{3-6}$  Because silicon has small spin-orbit coupling and an abundant isotope with zero nuclear spin, electron spins in silicon quantum dots have been predicted to have extremely long coherence times,$^{7, 8}$  a large advantage for spin-based quantum computing and for spintronics applications. These features have motivated efforts to develop quantum dots in silicon using a wide variety of confinement techniques.$^{9-18}$
 
Here we report the achievement of a single-electron quantum dot in a Si/SiGe modulation-doped heterostructure, in which an integrated charge-sensing quantum point contact$^{11, 19-23}$ is used to monitor electron transitions in and out of the dot and to verify the electron number.  Analogous single-electron quantum dots in GaAs/AlGaAs heterostructures have been used to form spin qubits -- quantum dots with spin states that can be manipulated and measured.$^{24-27}$
 
To achieve single-electron quantum dots in Si/SiGe heterostructures, one must overcome complications that do not arise in GaAs/AlGaAs heterostructures, including: (1) smaller Schottky barriers, leading to difficulty in the fabrication of low-leakage gates, (2) the need to implement strain management in Si/SiGe heterostructures, leading to disorder in the form of dislocations, mosaic tilt, and surface roughness, and (3) the larger effective mass of carriers in Si compared to GaAs, which decreases the tunneling rate through otherwise equivalent barriers to the leads.  Further, mobility in Si/SiGe is typically smaller than in III-V systems.  Our work builds on much recent progress in overcoming many of these issues in Si/SiGe, including the fabrication of gated quantum dots,$^{10-14}$ the observation of the Kondo and Fano effects in such a dot,$^{15}$ and the demonstration of transport through spin-channels in Si/SiGe double dots.$^{28}$  
 
The quantum dot used in this work was formed in a two-dimensional electron gas located 60 nm below the surface in a Si/SiGe heterostructure containing a Si quantum well.  Details of the sample can be found in reference.$^{28}$ The sample was illuminated for 20 seconds while at a temperature of 4.2 Kelvin at the beginning of the experiment before cooling the dilution refrigerator to base temperature, in order to decrease the resistance of the Ohmic contacts. The results we report below depend critically on the ability to apply large gate voltages without causing leakage currents. The quantum dot was formed on a mesa 10 microns wide by 20 microns long. The Schottky gates were formed by Pd deposition immediately following the removal of the native oxide by brief immersion in hydrofluoric acid.  The resulting Schottky gates supported applied voltages as large as -3.25 V relative to the electron gas.
  
\begin{figure}
\includegraphics{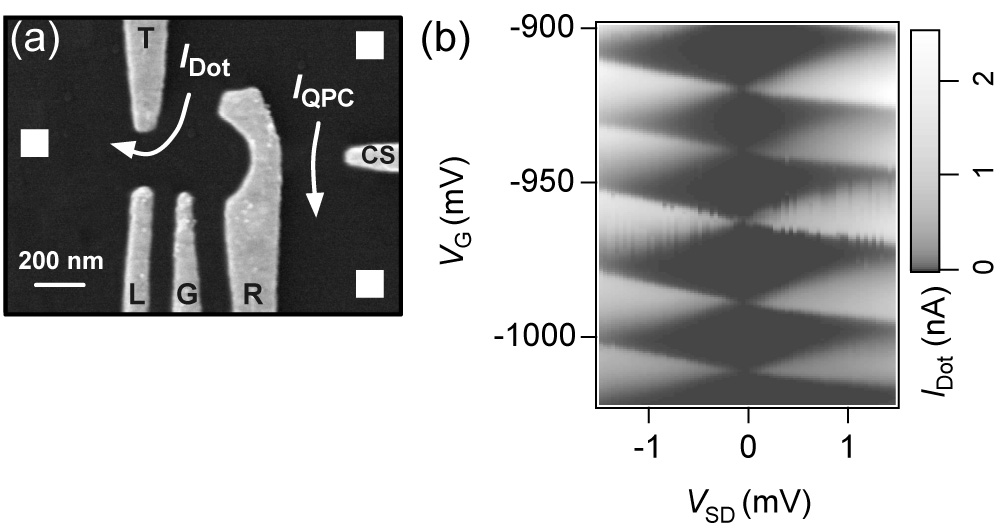}
\caption{(a) Scanning electron micrograph of a device with a design identical to the one used in this experiment.  Ohmic contact positions are indicated schematically as white squares, and the two current paths, through the dot and through the quantum point contact, are indicated schematically by white arrows.  (b) Gray-scale plot of the current through the dot as a function of the applied voltage across the dot and the gate voltage $V_\text{G}$.  The data were acquired by sweeping $V_\text{G}$ from more negative to less negative voltage for each value of $V_{\text{SD}}$.}
\end{figure}
  
A scanning electron micrograph of the top-gate design is shown in Fig. 1(a).  Negative voltages applied to the top gates deplete the underlying electron gas, forming both a single quantum dot defined by gates top (T), left (L), right (R), and the plunger gate (G), and an integrated charge-sensing quantum point contact (QPC) formed by the charge sensor gate (CS) and gate R.  Ohmic contacts to the 2DEG (shown schematically on the micrograph by superimposed white squares) were fabricated by evaporation of an Au:Sb (1$\%$) alloy with subsequent annealing at 550$^{\circ}$C.  A dc bias voltage across the top pair or the right pair of these Ohmic contacts causes current to flow through the quantum dot ($I_{\text{Dot}}$) or through the QPC ($I_\text{{QPC}}$) respectively.  Throughout this paper, the voltage on gate G ($V_\text{G}$) is used to control the number of electrons in the dot, and for the data presented here the measured electron temperature during the experiment was $30 \pm 20$ mK.

Figure 1(b) shows a Coulomb diamond plot of the source-drain current as a function of $V_\text{G}$ and the source-drain bias ($V_{\text{SD}}$).  Based on the electron counting discussed below, we estimate that the electron number in the regime of Fig. 1(b) is approximately 30.  As $V_\text{G}$ is made more negative, electrons are pushed off the dot one by one.  However, more negative $V_\text{G}$ also increases the potential barriers between the dot and the source and drain reservoirs, reducing the tunnel coupling between the leads and the dot.  This reduced tunnel coupling is visible in Fig. 2(b), where the Coulomb peaks decrease in height as $V_\text{G}$ is made more negative, until they are below the noise floor of the current measurement, which was 70 fA for the data presented here.  At this point, no further transitions in electron number can be monitored using direct current through the dot, and the introduction of a charge-sensing technique using the coupled QPC is essential.  It is interesting to note that the larger effective mass in Si (0.19 m$_{\text{e}}$) compared with GaAs (0.067 m$_{\text{e}}$) decreases the transparency of the tunnel barriers in Si as opposed to GaAs for the same electrostatic barrier shape and height. This may be one of the reasons that past measurements of Coulomb blockade in Si/SiGe have shown a relatively small number of Coulomb peaks.$^{15}$  For this reason, we focus in this paper on charge-sensing to confirm single-electron occupation.

\begin{figure}
\includegraphics{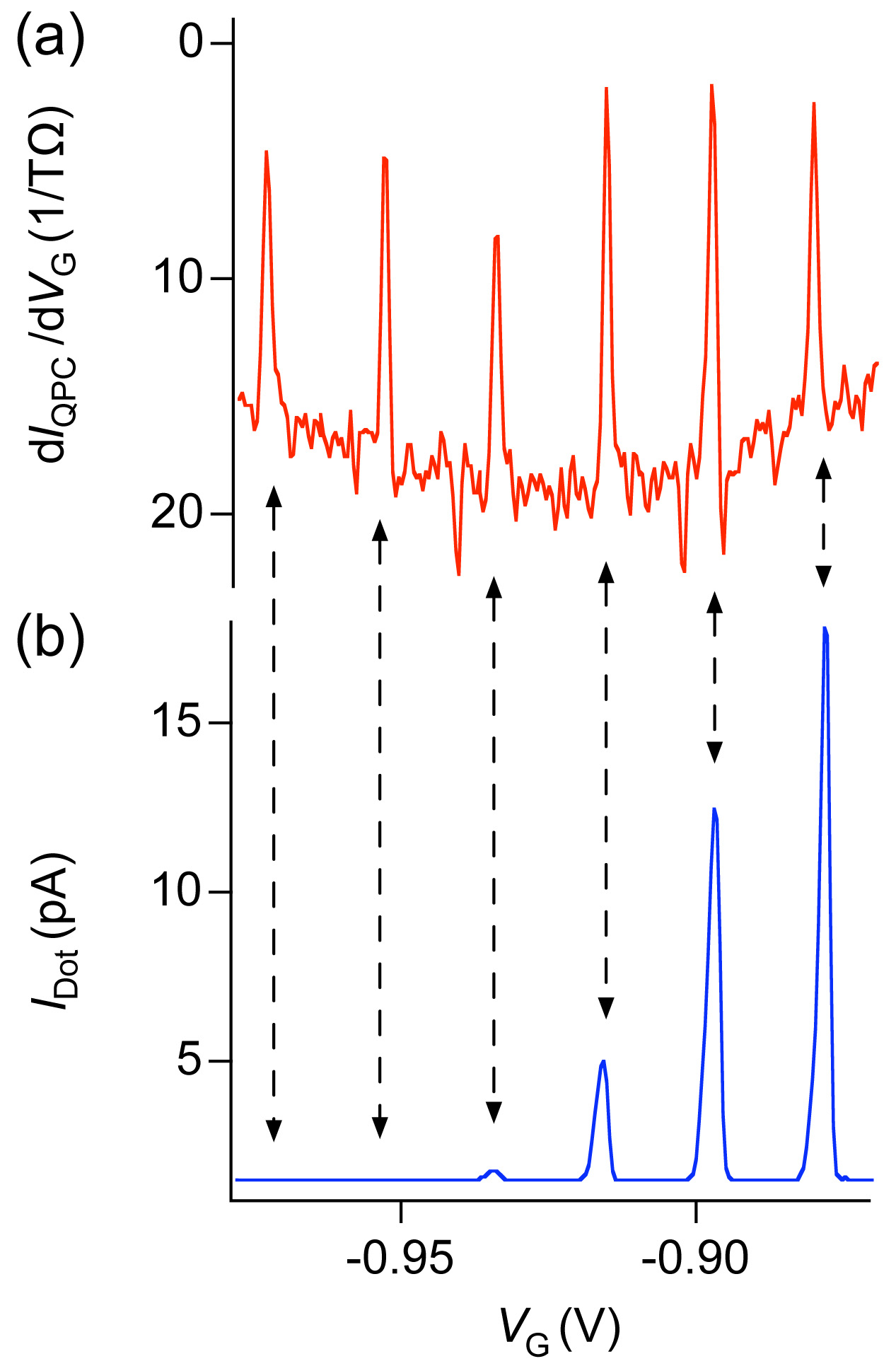}
\caption{(a) The derivative of the quantum point contact current with respect to the gate voltage d$I_{\text{QPC}}$/d$V_\text{G}$ as a function of the gate voltage $V_\text{G}$.  The peaks correspond to changes in the number of electrons in the dot.  (b) The current through the quantum dot as a function of the gate voltage $V_\text{G}$.  The peaks in (a) are well aligned with those in (b), indicating that the charge-sensing quantum point contact and the Coulomb blockade peaks in transport through the dot correspond to the same quantum dot charging phenomena.}
\end{figure}
  
Applying a negative voltage to gate CS, in combination with the effect of gate R, forms a QPC in close proximity to the quantum dot.  By precisely tuning the gate voltage $V_{\text{CS}}$, the conductance of the QPC can be fixed on a steep transition in the pinch-off curve.  In this configuration, the QPC functions as a sensitive electrometer for the neighboring quantum dot, because changes in the electron occupation of the dot result in measurable shifts in the QPC pinch-off curve.  Numerically differentiating $I_{\text{QPC}}$ with respect to $V_\text{G}$ turns these discrete shifts into peaks, and such a differentiated curve is plotted in Fig. 2(a).  The horizontal axes for the two plots are identical, and the data for each plot were acquired sequentially.  There is a clear correspondence between the peaks in the two curves, demonstrating that the QPC functions as a reliable detector of charge transitions in the quantum dot.  Importantly, this sensitivity is preserved even when transport through the dot is not measurable, as shown in Fig. 2.
	
The QPC is most sensitive to charge transitions in the quantum dot when its conductance varies rapidly as a function of gate voltage -- and hence also as a function of the charge on the dot.  However, changing $V_\text{G}$ to remove electrons from the dot also changes the potential of the coupled QPC.  The result is that, for a particular value of $V_{\text{CS}}$, there is a finite range over which $V_\text{G}$ can vary for which the QPC is sensitive to charge transitions on the dot.  Outside this range, the slope of the QPC conductance, which determines the sensitivity to charge transitions in the quantum dot, is too small to allow charge-sensing of single electrons.  In our system, transitions cannot be detected when d$I_{\text{QPC}}$/d$V_{\text{G}}$ is below 1 (T$\Omega$)$^{-1}$.  This provides an effective operational range of approximately 300 mV in $V_\text{G}$.  When the dot contains of order 30 electrons, this range is large enough to observe many charge transitions in the dot, because the spacing between the transitions is relatively small ($\sim$22 mV).  In the few electron regime, however, the spacing between transitions is larger and this range is not sufficient to observe more than three transitions with confidence.  Nonetheless, a large dynamic range can still be obtained by compensating the effect of $V_\text{G}$ on the QPC by changing  $V_{\text{CS}}$ in the opposite sense, keeping the QPC in the most sensitive operating point.
	
\begin{figure}
\includegraphics{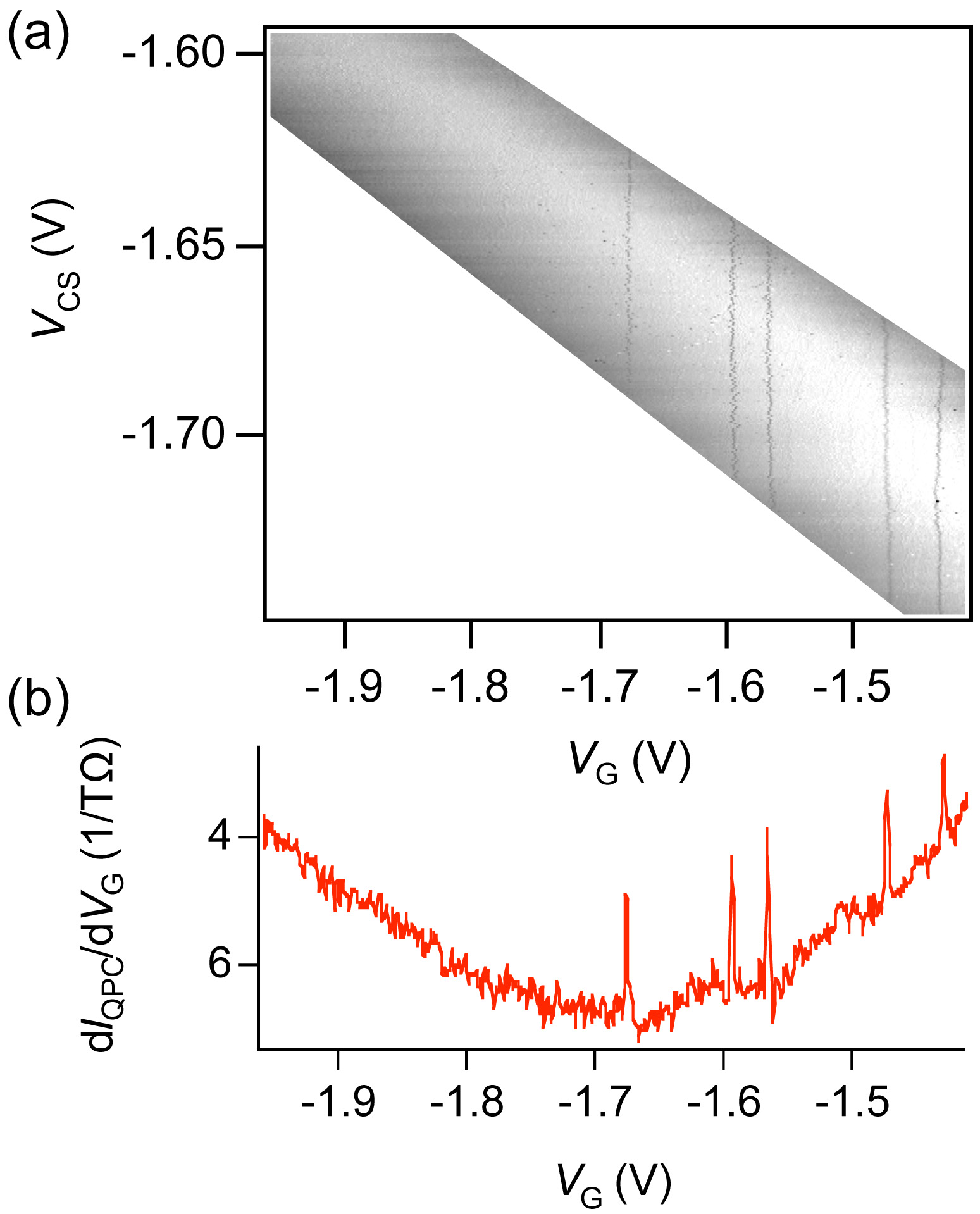}
\caption{(a) Gray-scale plot of the current through the charge-sensing quantum point contact.  The dark vertical lines correspond to changes in the quantum dot electron occupation.  No further transitions occur for $V_\text{G}$ $<$ -1.68 V, indicating that the quantum dot is empty of electrons in this 
regime.  (b) An average of 7 line-cuts taken diagonally down the sensitive slice in part (a).  The sharp peaks correspond to changes in the electron occupation of the dot.}
\end{figure}
  
An example of this type of compensation is shown in Fig. 3(a).  The voltage on gate G is swept through a range much larger than that corresponding to the sensitive region of the QPC.  By changing $V_{\text{CS}}$, high sensitivity is maintained across the entire range of $V_\text{G}$, so that many charge transitions can be monitored on a single image plot.  These charge transitions appear as the dark vertical lines in Fig. 3(a).  The spacing in gate voltage between the peaks is not uniform, as is expected for a dot with very few electrons.  The dot is empty of electrons for the most negative values of $V_\text{G}$, as indicated by the absence of dark lines on the left half of the figure.  A rigid shift was applied to each horizontal line-scan in Fig. 3(a) to remove two effects.  First, before the shift is applied the cross-capacitance between gate CS and the quantum dot causes the vertical lines in Fig. 3(a) to slope to more negative $V_\text{G}$ for less negative $V_{\text{CS}}$ with a lever arm of 26$\%$.  In addition, random charge fluctuations cause shifts from line-scan to line-scan with RMS magnitude 3.9 mV, which can be compared to the average spacing of 62.1 mV between the peaks.

Figure 3(b) shows an average of 7 diagonal line cuts taken parallel to the sensitive slice in Fig. 3(a).  The charge transitions appear as 5 sharp peaks, corresponding to the removal of the last 5 electrons from the dot.  The sequence of peaks terminates at $V_\text{G}$ = -1.68V, indicating that beyond this value of the gate voltage the dot is empty of electrons.  It is possible to then reduce the magnitude of $V_\text{G}$, refilling the dot with a known number of electrons starting from zero, something we have done many times over the course of the measurements reported here.\\

The authors thank Lisa McGuire, Mark Friesen, and Robert Joynt for helpful discussions.  This work was supported by NSA and ARO under contract number W911NF-04-1-0389, by NSF under grant numbers DMR-0325634 and DMR-0520527, and by DOE under grant number DE-FG02-03ER46028.\\
\\
\scriptsize
$^1$L. P. Kouwenhoven, T. H. Oosterkamp, M. W. S. Danoesastro, M. Eto, D. G. Austing, T. Honda, and S. Tarucha, Science 278, 1788 (1997).\\
$^2$L. P. Kouwenhoven, C. M. Marcus, P. L. McEuen, S. Tarucha, R. M. Westervelt, and N. S. Wingreen, in Mesoscopic Electron Transport, edited by L. L. Sohn, L. P. Kouwenhoven and G. Schšn (Kluwer, 1997), Vol. 345, p. 105.\\
$^3$D. Loss and D. P. DiVincenzo, Physical Review A 57, 120 (1998).\\
$^4$B. E. Kane, Nature 393, 133 (1998).\\
$^5$M. Friesen, P. Rugheimer, D. E. Savage, M. G. Lagally, D. W. van der Weide, R. Joynt, and M. A. Eriksson, Physical Review B 67, 121301 (2003).\\
$^6$X. D. Hu, B. Koiller, and S. D. Sarma, Physical Review B 71, 235332 (2005).\\
$^7$R. de Sousa and S. Das Sarma, Phys. Rev. B 68, 115322 (2003).\\
$^8$C. Tahan, M. Friesen, and R. Joynt, Physical Review B 66, 035314 (2002).\\
$^9$L. P. Rokhinson, L. J. Guo, S. Y. Chou, and D. C. Tsui, Physical Review B 63, 035321 (2001).\\
$^{10}$L. J. Klein, K. A. Slinker, J. L. Truitt, S. Goswami, K. L. M. Lewis, S. N. Coppersmith, D. W. van der Weide, M. Friesen, R. H. Blick, D. E. Savage, M. G. Lagally, C. Tahan, R. Joynt, M. A. Eriksson, J. O. Chu, J. A. Ott, and P. M. Mooney, Applied Physics Letters 84, 4047 (2004).\\
$^{11}$M. R. Sakr, H. W. Jiang, E. Yablonovitch, and E. T. Croke, Applied Physics Letters 87, 223104 (2005).\\
$^{12}$K. A. Slinker, K. L. M. Lewis, C. C. Haselby, S. Goswami, L. J. Klein, J. O. Chu, S. N. Coppersmith, R. Joynt, R. H. Blick, M. Friesen, and M. A. Eriksson, New Journal of Physics 7, 246 (2005).\\
$^{13}$L. J. Klein, K. L. M. Lewis, K. A. Slinker, S. Goswami, D. W. van der Weide, R. H. Blick, P. M. Mooney, J. O. Chu, S. N. Coppersmith, M. Friesen, and M. A. Eriksson, Journal of Applied Physics 99, 023509 (2006).\\
$^{14}$T. Berer, D. Pachinger, G. Pillwein, M. Muhlberger, H. Lichtenberger, G. Brunthaler, and F. Schaffler, Applied Physics Letters 88, 162112 (2006).\\
$^{15}$L. J. Klein, D. E. Savage, and M. A. Eriksson, Applied Physics Letters 90, 033103 (2007).\\
$^{16}$H. W. Liu, T. Fujisawa, Y. Ono, H. Inokawa, A. Fujiwara, K. Takashina, and Y. Hirayama, arXiv:0707.3513\\
$^{17}$J. Gorman, D. G. Hasko, and D. A. Williams, Physical Review Letters 95, 090502 (2005).\\
$^{18}$S. J. Angus, A. J. Ferguson, A. S. Dzurak, and R. G. Clark, Nano Letters 7, 2051 (2007).\\
$^{19}$M. Field, C. G. Smith, M. Pepper, D. A. Ritchie, J. E. F. Frost, G. A. C. Jones, and D. G. Hasko, Physical Review Letters 70, 1311 (1993).\\
$^{20}$E. Buks, R. Schuster, M. Heiblum, D. Mahalu, and V. Umansky, Nature 391, 871 (1998).\\
$^{21}$J. M. Elzerman, R. Hanson, J. S. Greidanus, L. H. W. van Beveren, S. De Franceschi, L. M. K. Vandersypen, S. Tarucha, and L. P. Kouwenhoven, Physical Review B 67, 161308 (2003).\\
$^{22}$T. Fujisawa, T. Hayashi, R. Tomita, and Y. Hirayama, Science 312, 1634 (2006).\\
$^{23}$Y. Hu, H. O. H. Churchill, D. J. Reilly, J. Xiang, C. M. Lieber, and C. M. Marcus, Nature Nanotechnology 2, 622 (2007).\\
$^{24}$J. M. Elzerman, R. Hanson, L. H. W. van Beveren, B. Witkamp, L. M. K. Vandersypen, and L. P. Kouwenhoven, Nature 430, 431 (2004).\\
$^{25}$J. R. Petta, A. C. Johnson, J. M. Taylor, E. A. Laird, A. Yacoby, M. D. Lukin, C. M. Marcus, M. P. Hanson, and A. C. Gossard, Science 309, 2180 (2005).\\
$^{26}$F. H. L. Koppens, C. Buizert, K. J. Tielrooij, I. T. Vink, K. C. Nowack, T. Meunier, L. P. Kouwenhoven, and L. M. K. Vandersypen, Nature 442, 766 (2006).\\
$^{27}$R. Hanson, L. P. Kouwenhoven, J. R. Petta, S. Tarucha, and L. M. K. Vandersypen, Reviews of Modern Physics 79, 1217 (2007).\\
$^{28}$N. Shaji, C. B. Simmons, M. Thalakulam, L. J. Klein, H. Qin, H. Luo, D. E. Savage, M. G. Lagally, A. J. Rimberg, R. Joynt, M. Friesen, R. H. Blick, S. N. Coppersmith, and M. A. Eriksson, arXiv:0708.0794
\end{document}